# Electrical Readout of Spin Environments in Diamond for Quantum Sensing

Olga Rubinas[1,2], Michael Petrov[1], Emilie Bourgeois[1,2], Jaroslav Hruby[1,2], Akhil Kuriakose[3], Ottavia Jedrkiewicz[4], Milos Nesladek[1,2]

[1]*Institute for Material Research (IMO), Hasselt University, Wetenschapspark 1, Diepenbeek, Belgium*
[2]*IMOMEC, IMEC, Kapeldreef 75, Heverlee, Belgium*
[3]*IFN-CNR and Dipartimento di Scienza e Alta Tecnologia, Università dell'Insubria, Via Valleggio 11, 22100 Como, Italy*
[4]*IFN-CNR and Como Lake Institute of Photonics, Università dell'Insubria, Via Valleggio 11, 22100 Como, Italy*

Corresponding Author: Milos Nesladek (milos.nesladek@uhasselt.be)

**ABSTRACT**. Nitrogen–vacancy (NV) centres in diamond are a key platform for quantum sensing and quantum information, combining long coherence times with controllable spin–spin interactions. Most of current quantum algorithms rely on optical access, which limit device integration and applicability in opaque or miniaturized settings. Here we demonstrate an all-electrical approach, photocurrent double electron–electron resonance (PC-DEER), permitting exploiting local dipolar interactions between individual NV spin qubits or ensembles and nearby paramagnetic defects with sub-confocal resolution. PC-DEER extends photocurrent NV readout from single-spin to spin-bath control and coherent manipulation, enabling characterization of bath-induced noise and effective deployment of noise-reduction protocols. We resolve the signatures of substitutional nitrogen (P1) and NVH centers with reproducible contrast by using electrical signals. Our results establish a scalable, optical-free spin readout strategy that bridges fundamental studies of spin environments with deployable quantum technologies, advancing the integration of diamond-based sensors into solid-state quantum devices.

## 1. INTRODUCTION.

Nitrogen-vacancy (NV) centers in diamond are a leading platform for quantum sensing [1] and quantum information processing [2] based on their long room-temperature coherence times [3] and strong luminescent transitions combíed with laser polarisation and MW manipulation. Optically detected magnetic resonance (ODMR) [4] has enabled studies ranging from fundamental spin physics, to single-spin imaging to nanoscale magnetometry and others. However, reliance on optical access constrains scalability, device integration, and applicability in opaque or miniaturized environments. In addition, the optical readout of the spin states is limited by the confocal resolution, limiting access to local spin-spin correlation to a small number of NV electron spin qubits, or alternatively to larger NV ensembles. With the progress in developing diamond qubit gates for dipole-dipole coupled NVs, there is demand for a local spin readout and scalability for qubit arrays as well as studying of the spin interaction between the spin qubit and the surrounding spin bath locally. The paper addresses a step in this beyond the optical readout, further developing the Photocurrent NV spin detection[5] to enable characterization of electron dipole-dipole coupling in diamond quantum electronic devices. Besides the spin coupling, our method gives information about spin-active point defects in the diamond materials, which is important for optimization of diamond for quantum applications. ODMR DEER has been used for this type of characterization mainly, based purely on optical transitions. The extension of DEER to electrical readout will allow to look also to the charge state dynamics, as the point defect occupation directly influences the charge carrier recombination dynamics, visible in PC-DEER.

Photocurrent NV spin detection [5] which we originated, offers an attractive alternative, converting spin-dependent ionization into measurable electrical signals and opening a route to chip-scale quantum devices. Previous photocurrent studies in both single NVs or NV ensembles have demonstrated basic spin control protocols, including Rabi oscillations [6], Ramsey fringes [7], Hahn echoes [8], and nuclear spin coherence via electrically detected ENDOR [9]. However, no complex multi-pulse sequences involving interactions with the surrounding spin bath—such as Double Electron–Electron Resonance [10] (DEER)—have been realized with electrical readout yet.

Here, we report a robust, repeatable implementation of DEER spectroscopy in NV ensembles using photocurrent detection (PC-DEER), achieving stable contrast and coherent spin manipulation of the bath. This capability enables electrical detection and control of environmental spins, a key step toward sensing external targets such as molecules or surface-bound radicals without optical access [11] or for studying many-body interactions. To our knowledge, this is the first photocurrent-based platform to meet the fidelity and contrast requirements for advanced quantum sensing protocols, bridging fundamental spin physics with scalable, integrated device architectures. As the photocurrent is sensitive to the charge state of the defects, our method can be potentially used for studying charge- spin dynamics. Moreover, our result extends photocurrent readout from single-spin control to coherent bath manipulation, providing a pathway toward noise-reduction protocols and spin-environment engineering. By eliminating the requirement for optical access, PC-DEER paves the way for integrating diamond-based quantum sensors into chip-scale architectures and for deploying them in complex or opaque environments relevant to future quantum technologies.

## 2. METHODS

### A. DEER Protocol and Spectral Features

The Double Electron-Electron Resonance (DEER) protocol was implemented as an extension of the Hahn-echo sequence. A π pulse resonant with the target spin bath (e.g., P1 centers) was inserted during the NV center's evolution time. This pulse flips the spin state of nearby P1 centers, thereby altering their dipolar coupling to the NV spin and inducing a phase shift in the NV coherence. The resulting change manifests as a reduction in the NV echo amplitude, revealing information about the surrounding spin environment.

NV center's spin state manipulation was performed on the frequency 2652 MHz – "$f_2$" on the FIG 1a. In applied magnetic field P1 center's splitting (FIG 1b) was in range of 100 – 400 MHz. The value of external magnetic field 78.6 G was determined by ODMR spectrum (FIG 1c) and aligned along one of the four possible ⟨111⟩ NV orientations. P1 centers in diamond have an electron spin S = ½ and a nuclear spin I=1, giving rise to three allowed transitions with $\Delta m_S = \pm 1, \Delta m_I = 0$ (FIG 1d). Due to the Jahn-Teller effect, P1 centers adopt four distinct crystallographic orientations, resulting in 12 resonance lines in the ideal DEER spectrum. However, with the external magnetic field aligned along a single NV axis, the projections of the field on the P1 axes split into two distinct groups: one with maximum alignment and the others with equal but reduced projections.

This orientation-dependent Zeeman splitting leads to two sets of P1 transitions, producing six distinguishable resonances. However, the central transitions ($|m_S = -\frac{1}{2}, m_I = 0\rangle \to |m_S = \frac{1}{2}, m_I = 0\rangle$) from both orientation groups are spectrally indistinguishable due to broad linewidths and high defect concentration. As a result, five distinct resonance groups are typically observed in the DEER spectra (FIG 1e).

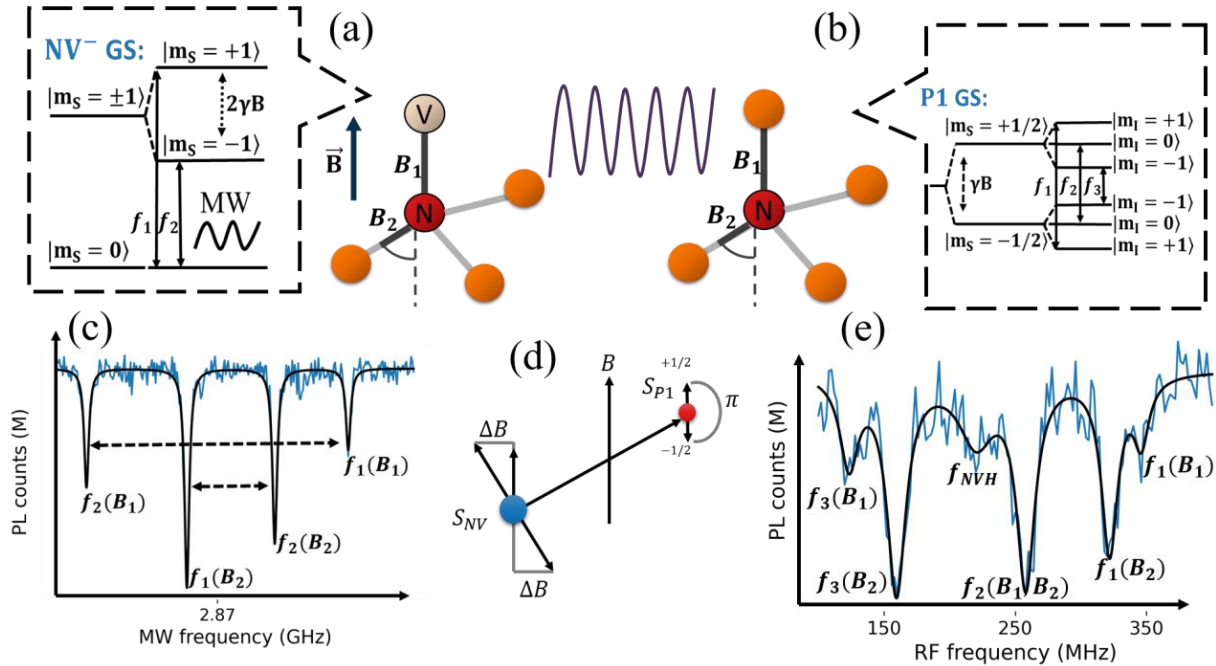

FIG 1. (a) Ground-state (GS) energy levels of the NV center, with electron spin transitions (arrows) labeled as transitions "$f_1$" and "$f_2$". Zeeman splitting is labeled by dashed arrow. On the lattice cell, there are two different magnetic field projections on the NV-axis. (b) GS energy level diagram of the P1 center, showing the three allowed electron spin transitions ("$f_1$", "$f_2$", and "$f_3$") by arrows, each is changed under different field projections. Zeeman splitting is labeled by dashed arrow. On the lattice cell, there are two different magnetic field projections on the P1-axis. (c) Representative NV ODMR spectrum with labeled transition frequencies and splitting due to two field projections. (d) Spin flip and navigated magnetic field from it scheme. (e) Optical detected DEER spectrum showing all detectable P1 center transitions $f_i$ (also divided into groups I–V as in main text) and NVH centers cluster.

Resonant transitions were combined into following groups for PC experiment: I – co-axis magnetic field orientated P1 centers ($f_3(B_1)$) with $|m_S = -\frac{1}{2}, m_I = -1\rangle \to |m_S = \frac{1}{2}, m_I = -1\rangle$, II – three non-co-axis P1 centers orientations ($f_3(B_2)$) with $|m_S = -\frac{1}{2}, m_I = -1\rangle \to |m_S = \frac{1}{2}, m_I = -1\rangle$, III – all orientations ($f_2(B_1, B_2)$) with $|m_S = -\frac{1}{2}, m_I = 0\rangle \to |m_S = \frac{1}{2}, m_I = 0\rangle$, IV – three non-co-axis P1 centers orientations ($f_1(B_2)$) with $|m_S = -\frac{1}{2}, m_I = +1\rangle \to |m_S = \frac{1}{2}, m_I = +1\rangle$ and V - co-axis P1 centers orientation ($f_1(B_1)$) with $|m_S = -\frac{1}{2}, m_I = +1\rangle \to |m_S = \frac{1}{2}, m_I = +1\rangle$. Because of very close position in spectrum of IV and V, we performed their PC-DEER measurements on one spectrum.

## B. Detection and Averaging

The PC signal arises from spin-dependent ionization of NV centers during laser illumination (FIG 2a). Spin transitions alter the population of metastable states, modulating the probability of ionization and resulting in measurable changes in current. This effect allows indirect readout of spin dynamics through electrical signals. This effect forms the basis of Photoelectric Detection of Magnetic Resonance (PDMR), an all-electrical alternative to the ODMR that enables spin state readout via changes in photocurrent rather than fluorescence (FIG 2b ,c).

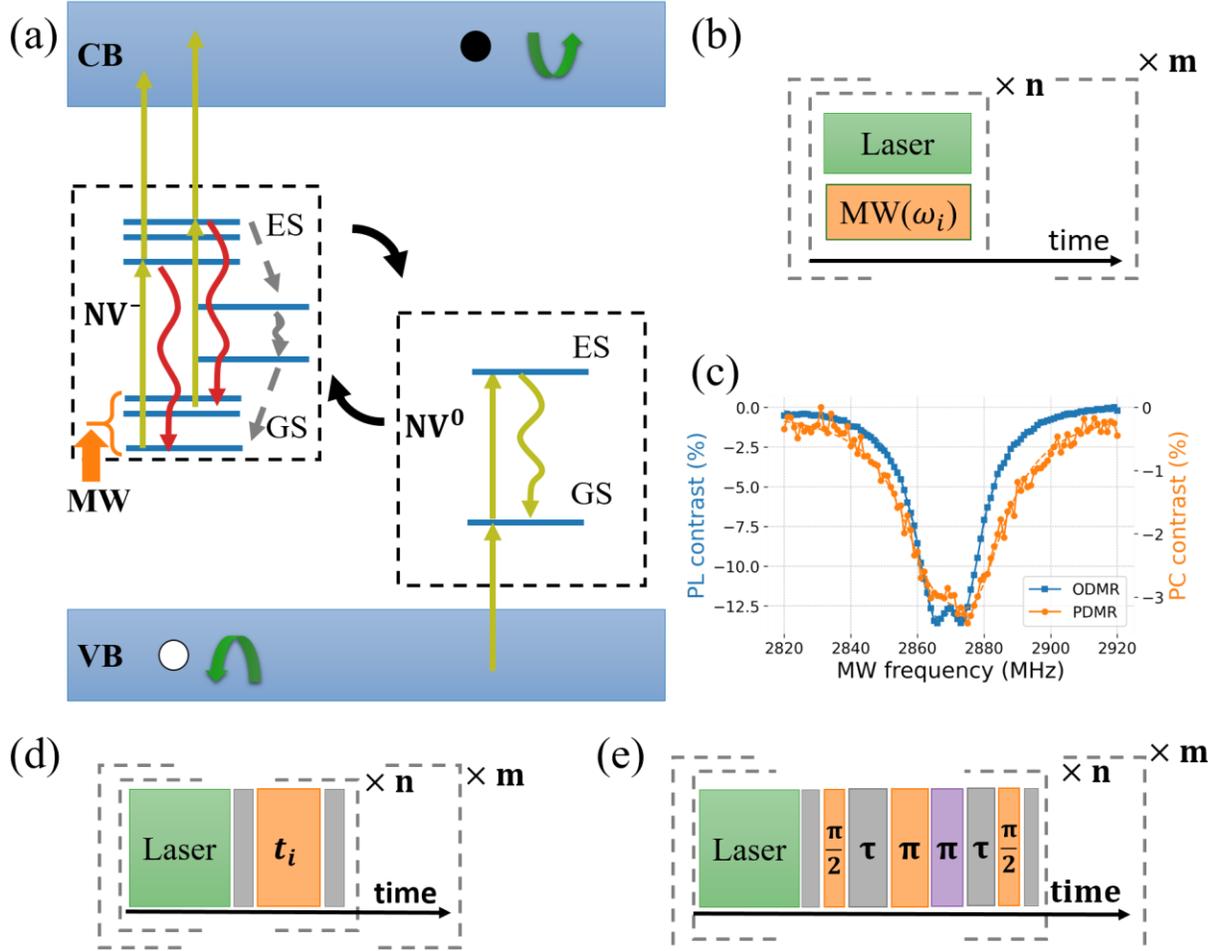

FIG 2. Principle of PC detection and PDMR measurement. (a) Energy level diagram illustrating the spin-dependent charge transfer mechanism in NV centers. (b) Schematic of the continuous-wave ODMR and PDMR experimental configuration. (c) Representative data showing both ODMR (optical) and PDMR (electrical) signals acquired at zero external magnetic field. Protocols for PC pulsed experiments: (d) Rabi oscillations of NV-center, $t_i$ −MW pulse duration. (e) DEER spectroscopy, $\tau$ −fixed free precession time of NV's spin, orange blocks mean MW $\pi/2$ and $\pi$ pulses, violet blocks mean RF pulse with varying frequency. Empty grey blocks are delays.

## 3. RESULTS

Before performing PC measurements, we conducted a full set of preparatory experiments using optical detection. A standard DEER spectrum was obtained for the P1 center ensemble under an external magnetic field of approximately 78.6 G (see Methods), with the field value determined via NV ODMR splitting. The DEER spectrum was divided into five principal groups, labeled I–V (see Methods), based on the positions of P1 transitions. Groups IV and V were combined in later measurements due to the close proximity of their resonances.

Each of the defined groups was individually targeted in our PC-based DEER protocol. As with standard optical experiments, the initial step involved measuring the PDMR signal at zero magnetic field, which yielded a contrast of 3.8% (FIG 2 in Methods).

We then performed ODMR splitting under an applied magnetic field, following the same approach used in optical calibration. A PDMR contrast of ~0.6% was observed for the NV center resonance at 2652 MHz (FIG 3a) (corresponding to a specific NV orientation).
Subsequently, we demonstrated PC-detected Rabi oscillations (FIG 3b) at an applied microwave power of $\approx 100$ mW before the PSB with a π-pulse duration of approximately 150 ns.

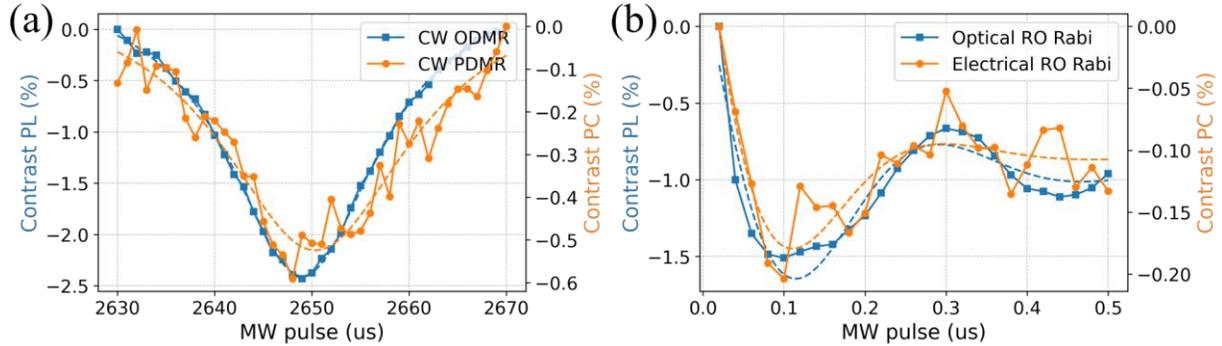

FIG 3. ODMR and PDMR on 2652 MHz resonance of one NV-orientation. (b) – Rabi oscillations of NV center with PL and PC readout observed at 20 mW of laser power.

For the final set of experiments—PC-DEER—the optimized parameters were: π-pulse (RF) applied to the P1 spin bath and NV spin free precession time $\tau_{fix}$. Validated by optical DEER oscillations π-pulse of p1 centers [13] was 60 ns. Due to optimization procedure, described in [14] NV spin free precession time $\tau_{fix}$ was chosen as 1400 ns.

To assess the performance of electrical detection in comparison with optical protocols, we recorded DEER spectra using three different readout configurations. The first spectrum, shown in the Methods section (FIG 1e), was acquired optically without point averaging ($n = 1$) and served primarily as a reference to identify expected resonance positions. The second spectrum was also obtained optically, but with repeated signal averaging at each point and full integration of the readout window. This configuration was executed in parallel with the PC-based measurements to ensure identical conditions. The third spectrum employed PC readout under the same averaging protocol as the second, with full integration of the signal and identical timing (FIG 4).

To compare the experimental data with theoretical predictions, the P1 center transitions (indicated by green lines in the spectrum) were calculated using its standard spin Hamiltonian:

$$H_{p1}/h = \frac{g_e \mu_B}{h} \vec{B} g \vec{S} + \frac{\mu_N}{h} \vec{B}\vec{I} + \vec{S}A\vec{I} + \vec{I}Q\vec{I}, \qquad (1)$$

where A and Q are isotopically symmetrical matrix with $A_\parallel = 114\ MHz$, $A_\perp = 81\ MHz$, $Q_\perp = -3.97\ MHz$. Because of symmetry quadrupole interaction is simplified as $Q_\perp S_z^2$.

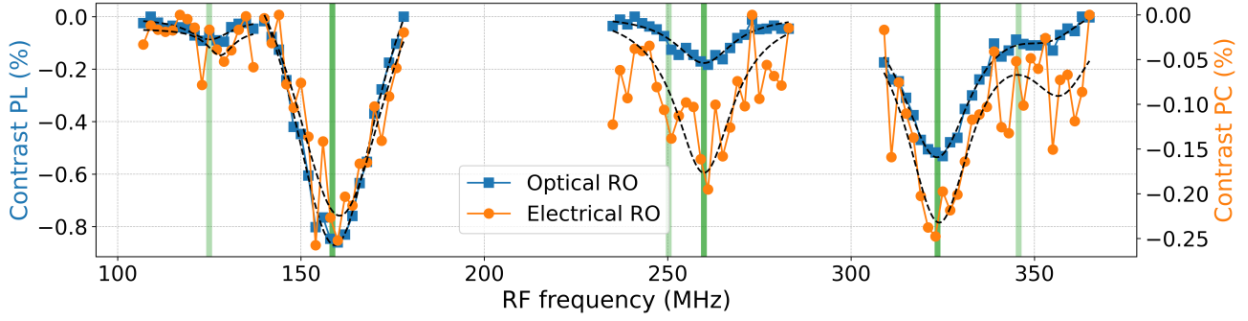

FIG 4. All parts (I – V) of p1 center DEER spectrum detected by optical and electrical readout. Green vertical lines correspond to allowed P1 transitions calculated from its Hamiltonian.

An additional broad resonance feature (FIG 5) was observed between 200 and 250 MHz ($f_{NVH}$ on the FIG 1e). This signal is likely attributable to NVH centers, a known defect in CVD-grown diamond [15]. Whilst presence of these defects was suggested in implanted CVD diamond by NV-DEER measurements, the presented measurements shows there presence also in in growth CVD N-doped material. These centers typically exhibit low DEER contrast even under optical detection, making electrical detection particularly challenging. Nonetheless, we were able to resolve this feature with measurable contrast using PC-DEER techniques

All allowed NVH transitions were also calculated from spin Hamiltonian parameters of NVH centers [16]:

$$H_{NVH^-}/h = \frac{g_e \mu_B}{h} \vec{B} g \vec{S} + \vec{S} A(H) \vec{I} + \vec{S} A(N) \vec{I}, \qquad (2)$$

where $A(N)_\parallel = 2.94\ MHz$, $A(N)_\perp = 3.1\ MHz$, $A(H)_\parallel = 13.69\ MHz$, $A(H)_\perp = -9.05\ MHz$, g = 2.0024. In our sample dominant isotope is 14N, so defect NVH- has S = ½ , I(H) = ½ and I(N) = 1. $H_{NVH^-}$ was also used for explanation of sides in some of our samples.

DEER spectra is broaden by high microwave power and high concentration of spin impurities. In this case, no hyperfine splitting with nitrogen in NVH is distinguished. Here we see only hydrogen (I = ½) splitting into two groups of resonances (violet lines on the FIG 5). For this group, we employed a 48 ns π-pulse, with the duration determined through prior optical calibration.

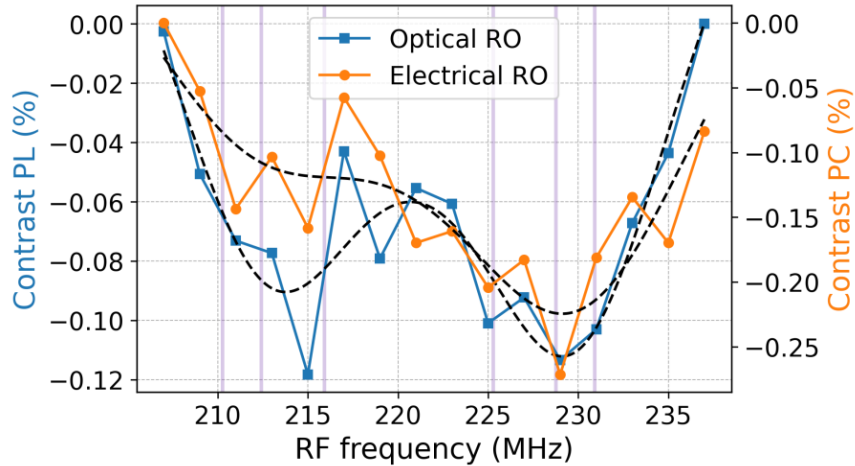

FIG 5. PL and PC DEER spectra for the possibly NVH center. Violet vertical lines correspond to allowed NVH transitions calculated from its Hamiltonian.

We additionally observed PC-detected Rabi oscillations of P1 centers within group II of the DEER spectrum (FIG 6). These oscillations appeared as periodic modulations in the DEER signal amplitude as a function of RF pulse duration, corresponding to coherent driving of the P1 spin transition (at 160 MHz for II group). The measurements were performed at a twice-lower RF power, resulting in a π-pulse duration of 80 ns. This demonstrates coherent control over the spin bath and confirms that its response can be monitored electrically via PC readout, widening up the pool of techniques for studying bath interactions.

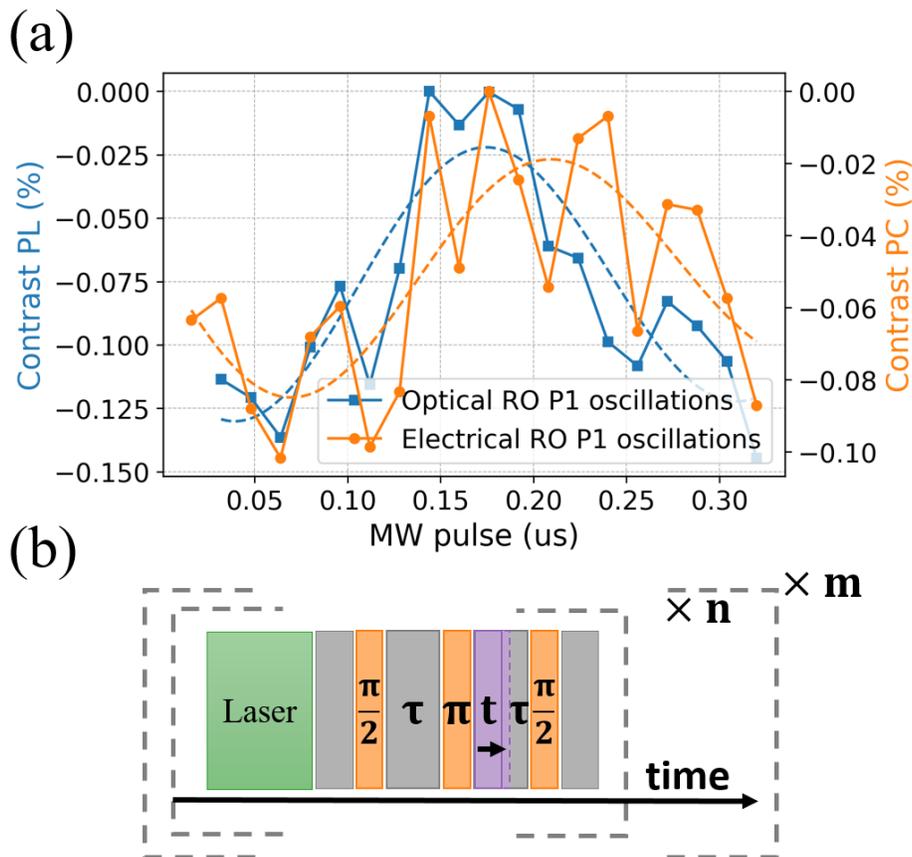

FIG 6. (a) PL and PC readout of P1 center's state $|m_S = -\frac{1}{2}, m_I = -1\rangle \rightarrow |m_S = \frac{1}{2}, m_I = -1\rangle$ oscillations for II group of resonances (160 MHz). (b) Pulsed scheme of P1 center's oscillations: grey blocks represent delay, $\tau$ is a fixed free precession time in DEER, orange blocks shows MW pulses, violet is RF pulse with varied duration, n is a point repetition; m is a full sequences repetitions.

### 4. DISCUSSION

All PC-based DEER measurements showed slightly lower contrast compared to their PL counterparts. Nonetheless, the demonstrated ability to detect the surrounding spin environment electrically represents a significant step towards exploring the active material characteristics that are essential for electrically detected spin-based devices and their miniaturization and integration such as for NV-based quantum sensors. This approach holds promise for the electrical detection of external spins, such as spin-labeled molecules [17] and surface-bound paramagnetic species [18], in compact device architectures [19, 20].

Future work will focus on pushing the sensitivity limits toward surface detection. Achieving this goal will require further optimization of diamond material properties—such as defect density, surface termination, and depth of NV centers—as well as the development of a new generation of high-efficiency electrodes tailored for spin-dependent PC detection. Whilst our technique was applied to micron-scale diamond devices, in follow up work we are extending the measurements to sub-100 nm scales where out technique can be used for quantum information purposes such as measuring quantum correlation in arrays of qubits and their interaction with the spin bath.

### 5. CONCLUSIONS

We have demonstrated PC-DEER spectroscopy in a NV ensemble within a CVD-grown diamond. This work represents the first successful realization of spin-bath driving and sensing via DEER using an all-electrical readout platform. Through systematic optimization of experimental parameters—including microwave pulse durations, NV spin precession time, laser power, and applied voltage—we achieved reproducible DEER contrast across five resonance groups attributed to P1 centers, and additionally resolved a weaker signal consistent with NVH-related transitions.

We also demonstrated coherent control of the spin bath by observing Rabi oscillations of P1 centers under PC readout, confirming that individual bath spin transitions can be selectively driven and detected electrically. This result establishes a clear pathway toward full quantum control of spin environments using PC techniques.

The ability to perform DEER spectroscopy and spin bath control electrically—despite the lower contrast compared to optical methods—marks a critical step toward scalable, chip-integrated NV-based quantum sensing.

These findings provide a foundation for advancing PC-based DEER techniques and extend their applicability to surface-bound spins and molecular systems in future device-integrated platforms.

**Competing interests:** Authors declare that they have no competing interests.

**Materials & Correspondence**: Correspondence and requests for materials should be addressed to Milos Nesladek (email: milos.nesladek@uhasselt.be) or Olga Rubinas (email: olga.rubinas@uhasselt.be).